\pdfoutput=1

\documentclass{PoS}
\usepackage{amsmath,amssymb}

\newcommand{\smallfrac}[2]{\mbox{\small ${\displaystyle \frac{#1}{#2}}$}}

\title{Lattice BRST without Neuberger 0/0 problem}

\ShortTitle{Lattice BRST without 0/0 problem}

\author{\speaker{Lorenz von Smekal} and Manon Bischoff\\
        Theoriezentrum, Institut f\"ur Kernphysik, Technische
        Universit\"at Darmstadt,\\ 
        Schlossgartestr. 2, 64289 Darmstadt, Germany\\ 
        E-mail: \email{lorenz.smekal@physik.tu-darmstadt.de}}

\abstract{We illustrate in a simple toy model how the methods of SUSY
  quantum mechanics and topological quantum field theory can be used
  for covariant gauge-fixing with unbroken BRST symmetry on a finite lattice.}

\FullConference{Xth Quark Confinement and the Hadron Spectrum\\
		 8-12 October 2012\\
		 TUM Campus Garching, Munich, Germany}

\begin{document}

\section{Introduction}
The covariant continuum formulation of gauge theories in terms of
local field systems relies on the existence of a well-defined and
unbroken Becchi-Rouet-Stora-Tyutin (BRST) symmetry. In particular, the
corresponding nilpotent BRST charge is needed to define the physical
subspace of the indefinite metric state-space of covariant gauge
theory in generalization of the Gupta-Bleuler condition in QED. This
is all very well understood in perturbation theory.
Beyond that, however, it is not so clear how to globally define such a
BRST charge in presence of the inevitable Gribov copies. On the
lattice, this problem is already present in the compact $U(1)$ gauge
theory. Already there, there is a perfect cancellation of
contributions from copies with even and odd numbers of negative
eigenvalues of the Faddeev-Popov operator ({\it i.e.}, even/odd Morse index)
to the measure in a standard BRST formulation. This cancellation is
the origin of the famous Neuberger $0/0$ problem of lattice BRST
\cite{Neuberger:1986vv}. Thus, this problem needs to be solved in the
compact $U(1)$ gauge theory already. The good news will be, however,
that a solution to the Neuberger $0/0$ problem in compact $U(1)$,
where it is a lattice artifact, is also suited for $SU(N)$ gauge
theories with little extra work. It is simply applied to the maximal
Abelian subgroup $U(1)^{N-1}$, the coset space $SU(N)/U(1)^{N-1}$ has no extra
$0/0$ problem \cite{vonSmekal:2008ws}. The corresponding lattice BRST for
gauge fixing the $SU(2)/U(1)$ coset space was explicitly constructed
already in Ref.~\cite{Schaden:1998hz}.   
   
After a short review of the standard procedure and its failure in the
next section, we will walk through the $U(1)$ problem in more detail in a simple
one-link model in Section 3. We explain the necessary extensions for $SU(2)$ in
Section 4 and provide our summary and outlook in Sec.~5. 
 
\section{Standard (double) BRST on the lattice}

The basic idea behind formulating (double) BRST on a finite lattice with the
methods of SUSY quantum mechanics is to formulate a topological Witten
model on the lattice gauge group whose partition function is to be used as
the gauge-fixing device. BRST  $s$ and anti-BRST  $\bar{s}$ variations
are thereby introduced as infinitesimal right multiplications of
the $SU(N)$ gauge group elements $g$ by Lie-algebra valued, anti-Hermitian ghost
and anti-ghost fields $c^\dagger = -  c$ and $\bar c^\dagger = - \bar c$,   
\begin{equation}
   s g \, = \, g \, X^a c^a \, = \, g c  \; , \qquad    
   \bar s g  \, = \, g \, X^a \bar c^a \, = \, g \bar c \; , 
\end{equation}
where $ [ X^a,X^b] = f^{abc} X^c$ with $\mathrm{tr}\, X^a X^b =
-\frac{1}{2}\delta^{ab}$.  
The partition function of this topological model must be independent
of the link variables $U_{ij}$ connecting nearest neighbor sites
$i\sim j$ which only enter via the gauge-fixing potential $V_U[g]$. In the
standard case, for example,
\begin{equation} 
 V_U[g] = - \sum_{i,j\sim i} \mathrm{Re} \, \mathrm{tr} \, U_{ij}^g \;
 ,\quad \mbox{where} \quad U_{ij}^g=g_i^\dagger U_{ij}g_j\; . 
\end{equation}
In terms of these, the BRST and anti-BRST
transformations then take the more familiar form,
\begin{equation}
sU_{ij}^g=-c_i U_{ij}^g+U_{ij}^gc_j \; , 
\qquad \bar{s}U_{ij}^g=-\bar{c}_i U_{ij}^g+U_{ij}^g\bar{c}_j \; .
\end{equation}
The (anti-)BRST transformations of ghost, anti-ghost and
Nakanishi-Lautrup fields $b^a$ act per site and are the same as in the
continuum \cite{ThierryMieg:1985yv}, 
\begin{equation} 
\begin{array}{lll}
sc^a = - \frac{1}{2}(c\times c)^a  &\qquad & \bar{s}\bar{c}^a =-
\frac{1}{2} (\bar{c}\times \bar{c})^a \\ 
s\bar{c}^a=b^a-\frac{1}{2}(\bar{c}\times c)^a &\qquad &
 \bar{s}c^a=-b^a-\frac{1}{2}(\bar{c}\times c)^a\\
sb^a=-\frac{1}{2}(c\times b)^a-\frac{1}{8}\big((c\times c)\times
\bar{c}\big)^a  &\qquad&  \bar{s}b^a=-\frac{1}{2}(\bar{c}\times
b)^a+\frac{1}{8}\big((\bar{c}\times \bar{c})\times c \big)^a
\end{array}
\end{equation}
where $(c\times c)^a\equiv f^{abc}c^bc^c$, etc. Ghost number and
Faddeev-Popov conjugation are part of a global $SL(2,\mathbb{R})$
symmetry in this extended double BRST formulation with gauge-fixing
action \cite{vonSmekal:2008en}, 
\begin{equation} 
S_{\mathrm{gf}} \, = - i \, s \bar s \, \Big( V_U[g] + i \frac{\xi}{2}  
\sum_i  \bar c_i c_i \Big) \,  
= \, 
\sum_{i}\Big(ib_iF_i[U^g]+i\bar{c}_iM_i[U^g,c]+\frac{\xi}{2}b_ib_i+\frac{\xi}{8}(\bar{c}_i\times  c_i)^2\Big) \, .
\label{S_GF}
\end{equation}
Here we dropped the color indices for brevity. The Faddeev-Popov operator $M $
is the Hessian of the Morse potential $V_U[g]$, and as such it is Hermitian
for all $\xi$.  In Landau gauge, $\xi=0$, the
$b$-fields establish $F_i[U^g] = 0$ and the corresponding gauge-fixing
partition function evaluates to
 \begin{equation}
Z_\mathrm{gf}=\int d[g,b,\bar{c},c] \exp\{-S_\mathrm{gf}\}\stackrel{\xi=0}{=}
\sum_{\mathrm{copies}}\frac{\text{det}\, M_F}{|\text{det}\, M_F|} \, .
\label{Z_GF}
\end{equation}
This is the sign-weighted sum over all Gribov copies whose vanishing
causes the $0/0$ problem of lattice BRST upon inserting $Z_\mathrm{gf}$
into the unfixed partition function $Z=\int d[U]\exp\{-S[U]\}$ of the 
gauge theory on the lattice. To see this explicitly one introduces
with Neuberger a parameter $t$, 
\begin{equation} 
S_{\mathrm{gf}}(t) \, = - i \, s \bar s \, \Big(\, t\, V_U[g] + i
\frac{\xi}{2}   
\sum_i \bar c_i c_i \Big) \, ,  \quad \mbox{such that} \quad \frac{d}{dt}
Z_\mathrm{gf}(t) = 0 \; ,
\end{equation}
just as $Z_\mathrm{gf}$ is independent of $\xi$ because both terms in
the action are separately BRST exact. Moreover observing that
$Z_\mathrm{gf}(t=0)=0$ then establishes the $0/0$ problem. The reason
for this is that $Z_\mathrm{gf} $ computes the Euler characteristic
$\chi$  of the lattice gauge group which is zero,
\begin{equation}
Z_\mathrm{gf}= \chi(SU(N)^{\times \# \mathrm{sites}})=\chi(SU(N))^{\#
  \mathrm{sites}}  , \quad \chi(SU(N)) =
\chi(S^3) \chi(S^5) \cdots \chi(S^{2N-1}) = 0 \, ,
\end{equation}
because of the $N-1$ odd spheres that make up the group manifold.   
And just as this zero was obtained for $\xi = 0$  in (\ref{Z_GF}) 
via the Poincar\'e-Hopf theorem \cite{Birmingham:1991ty}, here it follows from
one Gauss-Bonnet integral expression for $\chi(SU(N))$ per site on the
lattice \cite{vonSmekal:2008en}, 
\begin{equation}
Z_\mathrm{gf}(t=0) =\int
d[g,b,\bar{c},c]\; \exp\Big\{-\sum_i\Big(\frac{\xi}{2}b_ib_i+\frac{\xi}{8}(\bar{c}_i\times
    c_i)^2\Big)\Big\}\, . 
\end{equation}  
One possible remedy 
is to introduce a Curci-Ferrari mass term \cite{Kalloniatis:2005if} by
replacing 
$S_\mathrm{gf}$ with  
\begin{equation} 
S_{\mathrm{mgf}} =  - i \, (s \bar s +im^2) \, \Big(\, t\, V_U[g] + i
\frac{\xi}{2}   
\sum_i \bar c_i c_i \Big) \, .
\end{equation}
This decontracts the extended double BRST algebra. BRST
transformations are no-longer nilpotent, and the different Gribov copies get
reweighted by the explicit BRST breaking proportional to the
Curci-Ferrari mass parameter $m^2$. Instead of (\ref{Z_GF}) in Landau gauge one
then obtains
 \begin{equation}
Z_\mathrm{mgf}\stackrel{\xi=0}{=}
\sum_{\mathrm{copies}} \text{sign}\big(\text{det}\, M_F \big) \, 
 \exp\{-m^2 t V_U[g] \} \, ,
\label{Z_mGF}
\end{equation}
which lifts the cancellation of Gribov copies at the price of
unitarity violations in the corresponding continuum theory at finite $m^2$.
It also regulates the $t=0$ limit, {\it e.g.}, explicitly for $SU(3)$
\cite{vonSmekal:2008en},      
\begin{equation}
Z_\mathrm{mgf}(t=0) 
\propto (\xi m^4)^{\#\mathrm{sites}} \Big(1+ 4\xi
  m^4+\frac{64}{15}(\xi m^4)^2+\frac{64}{45}(\xi
  m^4)^3\Big)^{\#\mathrm{sites}} \, ,
\end{equation}  
and thus the $0/0$ problem. One might then hope to be able to restore
unitarity and compute observables in limit $m^2 \to 0$ from
l'Hospital's rule. Another way forward is to change the gauge-fixing
potential. For $SU(2)$ for example it was suggested in
\cite{vonSmekal:2007ns,vonSmekal:2008es} to replace    
\begin{equation}
V_U[g]=-\sum_{\mathrm{links}}\frac{1}{2}\text{tr} \, U^g \quad
\text{by} \qquad \widetilde
V_U[g]=-\sum_{\text{links}}\ln\big(1+\frac{1}{2}\text{tr}\, U^g \big)
\, . \label{gauge-pots}
\end{equation}
Near the identity they are essentially the same, so they are
equivalent, perturbatively, and they both lead to the same continuum
formulation. $\widetilde V_U$ is singular, however, whenever a gauge
orbit passes through the south pole.      
This amounts to setting up a Witten model, but instead of $\chi(S^3)=0$ for
$SU(2)$ based on 
\begin{equation}
\chi(\mathbb{R}P^2)=\frac{1}{2}\,\chi(S^2)=1\, ,
\end{equation}
for the $SU(2)/U(1)$ coset. The difference between the two lies in the
way the diagonal $U(1)$ subgroup is treated. We discuss this in a
toy-model next.

\section{One-link model for compact $U(1)$}

In order to illustrate the close connection to SUSY quantum mechanics
we consider the simplest case of a single compact $U(1)$ degree of
freedom $\varphi $ corresponding to a one-link model
\cite{Testa:1998az,Kalloniatis:2005if}. In addition we use a single 
angle $\theta \in (-\pi,\pi]$ for a non-periodic gauge transformation 
$\varphi^\theta \equiv \varphi-\theta$ which we furthermore allow to
depend on a fictitious time $\tau \in S^1$. With periodic boundary
conditions for all gauge dof's $\theta $, $b$ and Grassmann $\bar c$, $c$, we
can then write down an action for the corresponding toy model in 
SUSY quantum mechanics \cite{Birmingham:1991ty}, 
\begin{equation}
S_\mathrm{gf} 
=\int_0^1 d\tau \left[i\Big(\dot{\theta}+t\frac{\partial V_\varphi}{\partial
    \theta}\Big)b+\frac{\xi}{2}b^2 +
  \bar{c}\Big(\frac{d}{d\tau}+t\frac{\partial^2V_\varphi}{\partial 
    \theta^2}\Big)c \right] \, . \label{toy-model}
\end{equation}
The corresponding partition function is independent of gauge parameters
$t$, $\xi$ and link angle $\varphi$, and it is semiclassically
exact. If the height function on the circle is used as the Morse  
potential, $V_\varphi(\theta ) = 1 -  \cos(\varphi-\theta)$
for the standard lattice Landau gauge, one thus 
readily confirms that  
 \begin{equation}
 Z_\mathrm{gf} = \chi(S^1)=\frac{1}{|t|} \, \big(t-t \big) = 0 \, .
\end{equation}
The same will be true for any continuous periodic potential $V$ with
isolated zeroes on the circle. The link angle is inessential here and
will also be dropped in the following. We need to introduce a singularity in
$V$ to allow odd numbers of critical points and thus 
avoid this topological obstruction. Before we proceed, we note that
after redefining $t \to \xi t$ and then rescaling $\xi\tau \to \tau$,
\begin{equation}
S_\mathrm{gf} 
=\int_0^\xi d\tau \left[i\Big(\dot{\theta}+t\frac{\partial V}{\partial
    \theta}\Big)b+\frac{1}{2}b^2 + \bar{c}\Big(\frac{d}{d\tau}+t\frac{\partial^2V}{\partial
    \theta^2}\Big)c \right] \, .  \label{S_GFresc}
\end{equation}
This form allows to identify Landau gauge as the ``high temperature''
limit $\xi \to 0$ of a Witten model in which all non-constant modes
decouple, and in which the $\tau$-dependence is thus gone again. The
gauge-fixing partition function is given by the path integral 
representation of the Witten index $ \mathcal{W} = \Delta(\xi) $ of
the model in heat-kernel regularization,       
\begin{equation}
Z_\mathrm{gf} = \int \mathcal{D}[\theta,b,\bar{c},c]\, \exp\{-S_\mathrm{gf}\}
= \Delta(\xi) = \text{Tr}\big((-1)^{\mathcal{F}} e^{-\xi H}\big)
= \text{Tr}\big(e^{-\xi H_-}\big)-\text{Tr}\big(e^{-\xi H_+}\big) \,
. \label{WI-PI}
\end{equation}
Using $\bar c = c^\dagger$ for the Grassmann ghost  in the operator
language BRST and anti-BRST charges are identified with the
($N=2$) complex supercharges $Q$, $Q^\dagger$ and generalized ladder
operators $A$, $A^\dagger$,
\begin{equation}
\begin{split}
Q&=-\frac{i}{\sqrt{2}}bc^\dagger \equiv
A\sigma_+=\frac{1}{\sqrt{2}}\left(\frac{d}{d\theta}+\Phi(\theta)\right)\sigma_+
\, , \\
Q^\dagger &=\frac{i}{\sqrt{2}}b^\dagger c\equiv A^\dagger
\sigma_-=\frac{1}{\sqrt{2}}\left(-\frac{d}{d\theta}+\Phi(\theta)\right)\sigma_-
\, ,
\end{split} 
\end{equation}
 of the Witten model with SUSY potential $\Phi(\theta) = t V'(\theta)$
 and partner Hamiltonians
\begin{equation}
H_\pm = \frac{p^2}{2}+\frac{1}{2}\Phi^2(\theta)\pm\frac{1}{2}\Phi'(\theta)
    \equiv \frac{p^2}{2}+V_\pm(\theta)  \, .
\end{equation}
We can thus immediately write down the normalizable zero-energy ground state
solutions to 
\begin{equation}
\Big(\pm\frac{d}{d\theta}+\Phi(\theta)\Big) \, \phi_0^\mp(\theta)=0
\end{equation}
on the finite interval $\theta\in [-\pi,\pi]$  with periodic boundary
conditions $\phi(-\pi) = \phi(\pi)$. Whether the Witten index vanishes
or not only depends on what we choose for the (pre)potential
$V(\theta)$. 

Standard lattice Landau gauge thus corresponds to $V(\theta) = 1 -
\cos\theta $ with SUSY potential $\Phi(\theta) = t \sin\theta$, and
isospectral partner Hamiltonians with potentials $
V_\pm(\theta)=\frac{1}{2}(t^2\sin^2\theta\pm t\cos\theta)$. For any
value of $t$ they each have a normalizable zero-energy ground state
with wave function 
\begin{equation} 
\phi_0^\pm(\theta)=C\exp\{\pm t\cos\theta\} \, .
\end{equation}
As before, the Witten index counting the number of bosonic minus 
fermionic ground states is zero, $\Delta = n_- - n_+ = 1 - 1=0$.

\begin{figure}
\leftline{\hskip 4cm$\phi_0^-(\theta)$ \hskip 1.6cm $\phi_0^+(\theta)$
\hskip 0.3\linewidth $\phi_0^-(\theta)$ }
\vspace{-.6cm}
\centerline{\includegraphics[width=0.4\linewidth]{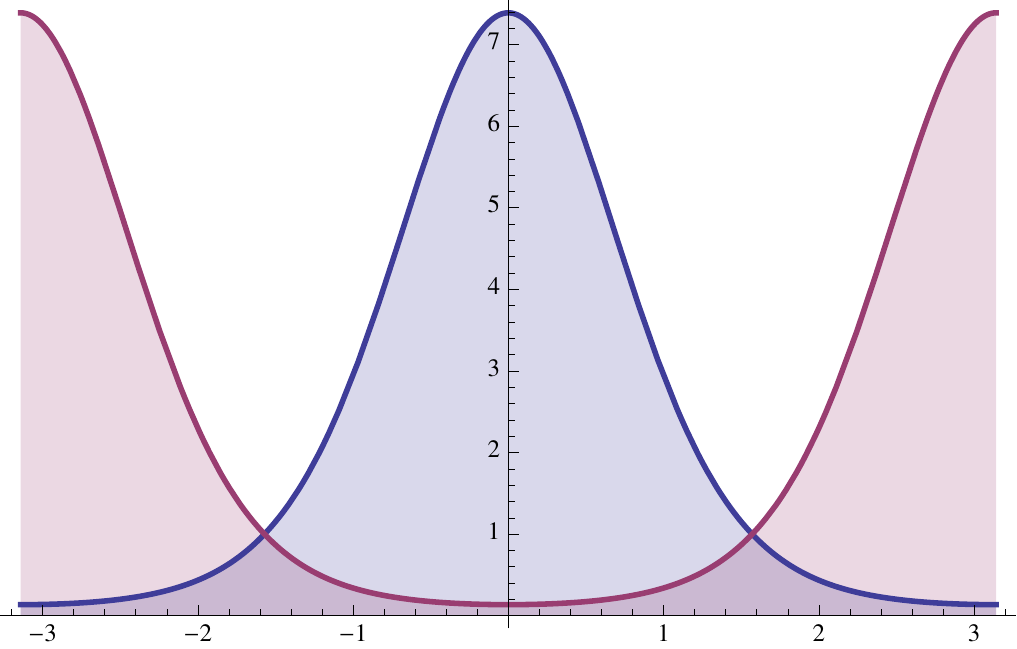}
\hspace{2cm}
\includegraphics[width=0.4\linewidth]{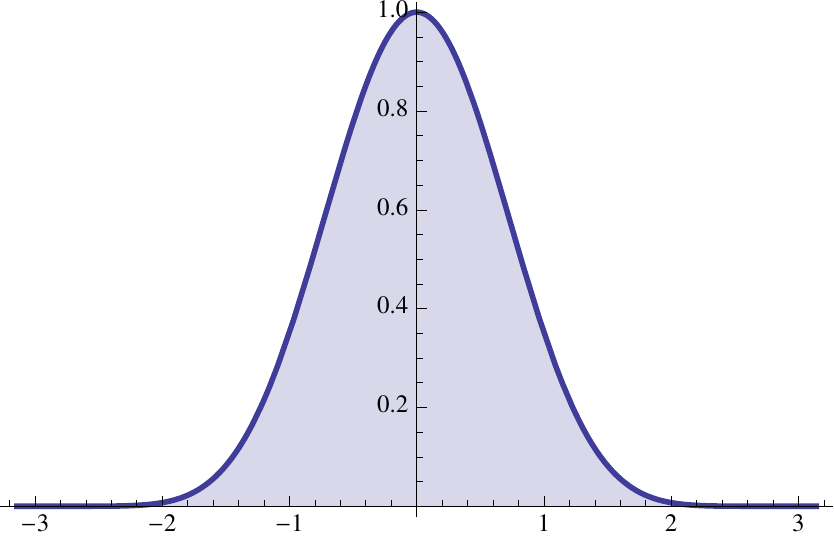}}
\vspace{-.3cm}
\leftline{\hskip .4cm $S$ \hskip 2.6 cm $N$ \hskip 2.6 cm $S$ \hskip 2.1cm 
$S$ \hskip 2.6cm $N$ \hskip 2.6 cm $S$}
\caption{Ground-state wave functions for standard (left) and modified
  (right) gauge-fixing potentials in the one-link $U(1)$ model
  (\protect\ref{toy-model}) with
  $t/\xi =2$. With the height function on the circle the
  bosonic ground state $\phi_0^-(\theta ) $ centered around the North Pole
  at $\theta=0$ is always accompanied by a fermionic 
  $\phi_0^+$ centered around the South Pole, $\theta= \pi$. In
  contrast, the P\"oschl-Teller potential has a unique bosonic ground
  state.}\label{fig1}  
\end{figure}

For the modified lattice Landau gauge, on the other hand, with
$V(\theta)=-2\ln\big((1+\cos\theta)/{2}\big)$, and SUSY potential 
$\Phi(\theta) = 2t\tan(\theta/2)$, the partner potentials 
\begin{equation}
V_\pm(\theta)=\frac{1}{8}\left(\frac{4t(4t\pm 1)}{\cos^2(\theta/{2})}-(4t)^2\right)   
\end{equation}
belong to the class of shape-invariant symmetric P\"oschl-Teller
potentials with good SUSY \cite{Junker:1996}, $\Delta = 1$, 
and a unique bosonic ground state with wave function
\begin{equation} 
\phi_0^-(\theta)=C \big( \cos(\theta/2) \big)^{4 |t|} \, .
\end{equation}
The ground-state wave functions for both cases are sketched in
Fig.~\ref{fig1}. As long as the Witten index $\mathcal W = \Delta(\xi)
= Z_\mathrm{gf}$ is non-zero, the corresponding SUSY on the gauge
group cannot break, and we will thus be guaranteed to have a well-defined
and unbroken BRST symmetry as well. The shape-invariant
P\"oschl-Teller oscillator for compact $U(1)$ has good SUSY, can be
solved exactly and it is straightforwardly generalized to a
one-dimensional chain.  

Before we continue, it will be useful to write the $U(1)$ toy model
in (\ref{toy-model}) in a manifestly coordinate and
metric independent form \cite{Birmingham:1991ty}. For a general $z =
f(\theta)$, the action in Eq.~(\ref{toy-model}) becomes, 
\begin{equation}
S_\mathrm{gf} =\int_0^1 d\tau \left[i
  \big(\dot{z}g(z)+tV'(z)\big)b+\frac{\xi}{2} g(z)b^2 + \bar{c}\Big( g(z)
  D_\tau+t \big(V''(z)-\Gamma(z) V'(z)\big)\Big) c\right] \, .
\label{S_z}
\end{equation}
A particularly convenient coordinate 
is given by stereographically
projecting the circle $S^1 \to \mathbb R$ via $z= 2 \tan(\theta/2)$
for the modified potential $V(z) = 2 \ln(1+z^2/4)$, with 
one-dimensional metric
\begin{equation}
g(z) = \frac{1}{(1+z^2/4)^2}\, , \quad \Gamma = \frac{d}{dz} \ln\sqrt g
    = \frac{-z/2}{1+z^2/4} \, ,
\label{one-d-met}
\end{equation} 
and covariant derivative $D_\tau = d/d\tau + \Gamma \dot z $. The fact
that $Z_\mathrm{gf} = \mathcal W = 1$, is then readily verified once more  
from the path integral representation (\ref{WI-PI}) with the simple Nicolai
map  $\eta \equiv \dot z/(1+z^2/4) + t z $.

\section{From compact $U(1)$ to $SU(2)$}

Since we intend to use the supersymmetric P\"oschl-Teller oscillator also 
for a $U(1)$ subgroup in $SU(2)$ we first focus on the
two-dimensional coset $SU(2)/U(1) \simeq S^2$ (per site). The Euler
characteristic of a two-dimensional compact manifold $\mathcal M$ is given
by the integral over its Gauss curvature $K = {R^{12}}_{12}$, the only
independent component of the Riemann curvature tensor $R_{ijkl}$ in 2
dimensions,
\begin{equation}
\chi(\mathcal{M})=\frac{1}{2\pi}\int_{\mathcal{M}}K\, dv \, .
\end{equation}
In particular,  $K= 1/R^2$ for a sphere $S_R^2$ of radius $R$, $dv = R^2
d\Omega $ and the 
Euler characteristic is of course independent of $R$,
\begin{equation}
\chi(S_R^2)=\chi(S^2) =\frac{1}{2\pi}\int_{S^2} d\Omega = 2 \, .
\end{equation}
A representation of this same Gauss-Bonnet formula which
however holds for compact manifolds without boundary of any even dimension $2n$
involves $2n$ pairs of  Grassmann variables $\bar c$, $c$, see 
 \cite{Birmingham:1991ty},
\begin{equation}
\chi(\mathcal M)=\int_{\mathcal M} d^{2n}x \int_{\mathbb R^{2n}} \frac{d^{2n}b}{(2\pi)^{2n}} \int
d[\bar{c},c]\, \exp\left\{-\frac{\xi}{2}b^ib^jg_{ij} + \frac{\xi}{4}
  R_{ijkl} \bar{c}^ic^k\bar{c}^jc^l \right\} \, .  \label{Gauss-Bonnet}
\end{equation}
For odd-dimensional manifolds the corresponding result is
automatically zero as it must because the exponential only produces
even powers of the Grassmann variables. 
For the even-dimensional spheres $S_R^{2n}$ it is straightforward to
verify that $\chi(S_R^{2n}) = 2 $ from this formula. This can be done
explicitly, {\it e.g.}, with again using stereographic coordinates
$\vec x \in \mathbb R^{2n}$ with $x=2R\tan(\theta/2)$, where $\theta
$ is the azimuthal angle, 
metric $ g_{ij}= (1+ x^2/(2R)^2)^{-2}\, \delta_{ij}$ and curvature
$R_{ijkl} = R^{-2}(g_{ik} g_{jl} - g_{il} g_{jk})$.  

For the projective space $\mathbb RP^{2} $ one either integrates only over one
hemisphere to the equator at $x=2R$ or simply divides the full
integral over $\vec x \in \mathbb R^{2}$ for $S^2$ by two as mentioned above. 
Since the Gauss-Bonnet integral for the Euler characteristic of a
sphere is independent of its radius $R$ we may as well integrate
$\chi(S_R^2) $ over
all $R$ with unit weight, $1= \int dR \, w(R)$, without changing the
result, {\it i.e.}, $\chi(S^2) = \int dR \, \chi(S_R^2) \, w(R)$. For this
normalized integral we use the Witten index $\mathcal W =
Z_\mathrm{gf} = 1$ of the P\"oschl-Teller oscillator with
$S_\mathrm{gf}$ from (\ref{S_z}) for $U(1)$ in the last section by
identifying the radius $R$ of the two-sphere as $R= z = 2\tan(\theta/2)$
where $\theta $ is now the azimuthal angle of $S^3$ for $SU(2)$.   

What we have then achieved is to integrate in our final
gauge-fixing partition function $Z_\mathrm{gf} $ for  $SU(2)$ 
over the whole group manifold, however with $\theta \in (-\pi,\pi)$
in the P\"oschl-Teller oscillator rather than $[0,\pi)$ for the usual
azimuthal angle of $S^3$. The remaining two $S^3$ coordinates $\vec x
$ then parametrize $\mathbb RP^2$ instead of $S^2$, hence we include a
factor $1/2$ (per site). The  partition function for the corresponding
one-link model for $SU(2)$ then explicitly becomes
\begin{equation} 
Z_\mathrm{gf}^{SU(2)}  = \smallfrac{1}{2}  \int \mathcal D[z, b_z, \bar
c_z, c_z] \mathcal D[x,b_x,\bar c_x, c_x ] \, \exp\{-
S_\mathrm{gf}^{(x)}[x,b_x,\bar c_x,c_x]  -
S_\mathrm{gf}^{(z)}[z,b_z,\bar c_z,c_z]\}\, ,
\end{equation}
with $S_\mathrm{gf}^{(z)}[z,b,\bar c ,c]$ for the P\"oschl-Teller
oscillator as given in (\ref{S_z}) and
\begin{equation} 
S_\mathrm{gf}^{(x)}[x,b,\bar c,c] = \int_0^1 d\tau  \, \Big[
i  g_{ij} \dot x^i b^j + \frac{\xi}{2} g_{ij} b^i b^j - \frac{\xi}{4}
R_{ijkl} \bar c^i c^k\bar c^j c^l + \bar c^i g_{ij} (D_\tau c)^j
\Big]    \, , \label{S-S2}
\end{equation}
with $i,j = 1,2$. Since we did not need to introduce a Morse potential
on $S^2$, the action in (\ref{S-S2}) is itself the $t\to 0$ limit of a
Witten model to compute $\mathcal W = \chi(S^2)$. In this limit, only
the constant modes contribute and the path integral reduces to the
corresponding Gauss-Bonnet integral (\ref{Gauss-Bonnet}). If we
however now change coordinates from $z= 2\tan(\theta/2) \in \mathbb R
$, or $\theta \in (-\pi,\pi)$, and $\vec x \in \mathbb RP^2$ to general
coordinates for $S^3$, the P\"oschl-Teller 
prepotential $V(\theta) $, here defined as
function of the class angle $\theta $ of $SU(2)$, will then depend on
all three of the new coordinates. For example, with stereographic
coordinates $\vec x \in \mathbb R^3$ with $x = 2\tan(\theta/2)$ for
$S^3$, the total gauge fixing action, the sum of
$S_\mathrm{gf}^{(x)}$ and  $S_\mathrm{gf}^{(z)}$, will be of the form
\begin{align} 
S_\mathrm{gf}^{(\mathrm{tot})} &= \int_0^1 d\tau  \, \Big[
i  \Big(\dot x^a g_{ab} + t \frac{\partial V}{\partial x^b} \Big)  b^b
+ \frac{\xi}{2} g_{ab} b^a b^b - \frac{\xi}{4} 
R_{abcd} \bar c^a c^c\bar c^b c^d \\ 
& \hskip 4cm + \bar c^a \Big( g_{ab} (D_\tau
c)^b + t \Big( \frac{\partial^2 V}{\partial x^a \partial x^b} -
\Gamma^c_{ab} \frac{\partial V}{\partial x^c} \Big) c^b \Big)   \nonumber
\Big]    \, , \label{S-S3}
\end{align}
where now $a,b,c = 1,2,3$.  It is important to remember here, however,
that the metric $g_{ab}$ is the transformed product metric for $S^2
\times S^1 $, an originally block-diagonal one consisting of a
$2\times 2$ metric $g_{ij}$  for $S_R^2$ and a one-dimensional metric as in
(\ref{one-d-met}) for $S^1$. Analogously, the curvature tensor here is
the correspondingly transformed $R_{ijkl} = R^{-2}(g_{ik} g_{jl} -
g_{il} g_{jk})$ with $i,j = 1,2$ for $S_R^2$. In particular, it is not
the curvature of $S^3$ or $SU(2)$ as in (\ref{S_GF}) with $R_{abcd} =
\frac{1}{2} \varepsilon_{abe} {\varepsilon^e}_{cd} $ which would
produce the Neuberger zero for $t\to0$ and with a bounded gauge-fixing
potential $V$.  

A rescaling of the time interval as in (\ref{S_GFresc}) but now from
$\tau \in [0,1]$ to $\tau \in [0,t]$ with $\xi \to t\xi$ shows that
also for $t\to 0 $ only constant modes survive, likewise. With one such set
of constant modes to be integrated per site $i$ on a lattice, where
the only coupling of neighboring sites comes from the gauge-fixing
potential, the covariant gauge-fixing action for $SU(2) $ without
$0/0$ problem becomes,
\begin{equation} 
S_\mathrm{gf} = \sum_i  \Big\{
i  \frac{\partial V}{\partial x_i^a}  b_i^a
+ \frac{\xi}{2} g_{ab} b_i^a b_i^b  - \frac{\xi}{4} 
R_{abcd} \bar c_i^a c_i^c\bar c_i^b c_i^d  \Big\}
 + \sum_{i,j}  \bar c_i^a \Big( \frac{\partial^2 V}{\partial x_i^a \partial x_j^b} -
\delta_{ij} \Gamma^c_{ab} \frac{\partial V}{\partial x_j^c} \Big) c_j^b  \, . \label{S-lat}
\end{equation}
This is essentially of the same form as in (\ref{S_GF}). The only
differences are in the $S^2\times S^1$ metric and curvature to be used here as
discussed above, and the singular gauge-fixing potential with $V \equiv
\widetilde V_U[g]$ from (\ref{gauge-pots}) in order to quantize one
P\"oschl-Teller oscillator for the $SU(2)$ class angle per site on the
lattice so that the corresponding
$ Z_\mathrm{gf} =  \chi(\mathbb RP^2)^{\#\mathrm{sites} } = 1$. 
The fact that the modified gauge-fixing potential 
must have a singularity thereby only 
means that contributions from a gauge orbit close to that singularity,
for $\widetilde V_U[g]$ in (\ref{gauge-pots}) at $U^g= - 1$,  are
exponentially suppressed.    

\section{Summary and Outlook}
 
Starting from a simple one-link model, we have described explicitly
how the problem of gauge-fixing on the lattice can be formulated in
terms of the Witten index in SUSY quantum mechanics. All excited states of
gauge and ghost degrees of freedom then cancel and the gauge-fixing partition
function is determined entirely by the zero-energy ground
states. Therefore, after gauge-fixing the contribution of each gauge
orbit is represented by the difference of bosonic and fermionic ground
states in the Witten model along this orbit. The BRST and anti-BRST
symmetries then correspond to the $N=2$ supercharges of the Witten
model, and as long as the Witten index is non-zero, one is guaranteed
to have a good SUSY and thus a well-defined and unbroken BRST symmetry.

This is all verified explicitly in our one-link model for $U(1)$
and $SU(2)$. The generalization to a one-dimensional chain is
straightforward. The P\"oschl-Teller oscillators for compact $U(1)$ 
only have bosonic ground states in higher dimensions as well
\cite{vonSmekal:2007ns}, there are thus no cancellations. Counting these
bosonic ground states in more than one dimension is a challenging
problem, however  \cite{Mehta:2009zv}. The generalization to $SU(N)$
is not entirely straightforward either and currently in progress.   

\bigskip

This work was supported by the Helmholtz International Center for FAIR
within the LOEWE program of the State of Hesse and the European
Commission, FP7-PEOPLE-2009-RG No. 249203.

\end{document}